\documentclass[conference]{IEEEtran}
\IEEEoverridecommandlockouts
\usepackage{cite}
\usepackage{amsmath,amssymb,amsfonts}
\usepackage{algorithmic}
\usepackage{graphicx}
\usepackage{adjustbox}
\usepackage{amssymb}  % For checkmark symbols
\usepackage{dblfloatfix}

\usepackage{amsmath}
\usepackage{colortbl}
\usepackage{xcolor}
\usepackage{textcomp}
\usepackage{booktabs} % For better table formatting
\usepackage{placeins}

\usepackage{xcolor}
\def\BibTeX{{\rm B\kern-.05em{\sc i\kern-.025em b}\kern-.08em
    T\kern-.1667em\lower.7ex\hbox{E}\kern-.125emX}}

\begin{document}

\title{GSAC: Leveraging Gaussian Splatting for Photorealistic Avatar
Creation with Unity Integration\\

\thanks{Identify applicable funding agency here. If none, delete this.}
}

\author{\IEEEauthorblockN{ Rendong Zhang}
\IEEEauthorblockA{\textit{Dept. of Computer Science } \\
\textit{Vanderbilt University}\\
Nashville, USA\\
rendong.zhang@vanderbilt.edu}
~\\
\and
\IEEEauthorblockN{ Alexandra Watkins }
\IEEEauthorblockA{\textit{Dept. of Mechanical Engineering} \\
\textit{Vanderbilt University}\\
Nashville, USA\\
alexandra.watkins@vanderbilt.edu}

~\\

\and
\IEEEauthorblockN{ Nilanjan Sarkar}
\IEEEauthorblockA{\textit{Dept. of Mechanical Engineering} \\
\textit{Vanderbilt University}\\
Nashville, USA\\
nilanjan.sarkar@vanderbilt.edu}
*Corresponding author
}

\maketitle

\begin{abstract}
Photorealistic avatars have become essential for immersive applications in virtual reality (VR) and augmented reality (AR), enabling lifelike interactions in areas such as training simulations, telemedicine, and virtual collaboration. These avatars bridge the gap between the physical and digital worlds, improving the user experience through realistic human representation. However, existing avatar creation techniques face significant challenges, including high costs, long creation times, and limited utility in virtual applications. Manual methods, such as MetaHuman, require extensive time and expertise, while automatic approaches, such as NeRF-based pipelines often lack efficiency, detailed facial expression fidelity, and are unable to be rendered at a speed sufficent for real-time applications. By involving several cutting-edge modern techniques, we introduce an end-to-end 3D Gaussian Splatting (3DGS) avatar creation pipeline that leverages monocular video input to create a scalable and efficient photorealistic avatar directly compatible with the Unity game engine. Our pipeline incorporates a novel Gaussian splatting technique with customized preprocessing that enables the user of "in the wild" monocular video capture, detailed facial expression reconstruction and embedding within a fully rigged avatar model. Additionally, we present a Unity-integrated Gaussian Splatting Avatar Editor, offering a user-friendly environment for VR/AR application development. Experimental results validate the effectiveness of our preprocessing pipeline in standardizing custom data for 3DGS training and demonstrate the versatility of Gaussian avatars in Unity, highlighting the scalability and practicality of our approach.
\end{abstract}

\begin{IEEEkeywords}
Photorealistic Avatar, Gaussian Splatting, Virtual Reality, Augmented Reality
\end{IEEEkeywords}

\section{Introduction}
We live in an era of rapid technological advancement, marked by groundbreaking innovations in artificial intelligence (AI), machine learning (ML), and immersive technologies like virtual reality (VR) and augmented reality (AR). Initially focused on entertainment, VR and AR have now expanded their influence to sectors such as education, healthcare, and industry, transforming how we connect, learn, and collaborate\cite{XandraEveryBody}. These technologies provide immersive, interactive experiences that redefine human interaction. However, at the core of these transformative applications lies a critical need: the natural and realistic representation of humans in virtual spaces.

Virtual avatars serve as the linchpin for realizing the full potential of VR and AR, offering a means to foster lifelike interactions and create engaging virtual environments. From training simulations to telemedicine, realistic avatars are essential to delivering meaningful, real-time experiences. Yet, achieving authenticity in human avatars—capturing both appearance and behavior—remains a significant challenge. This underscores the necessity for advanced methods to bridge the gap between the physical and virtual worlds, enabling realistic and impactful interactions in virtual environments.

Moreover, those avatars play a crucial role in enhancing the immersive and co-present nature of interactive VR and AR experiences. Immersion, the sense of deep engagement in a virtual environment, and co-presence, the feeling of sharing a space with others, are strengthened by photorealistic avatars \cite{PRAvatarIncreaseCo-Presence}, when avatars effectively convey nonverbal cues such as facial expressions, gestures, and eye movements. 
By bridging the physical and digital realms, photorealistic avatars elevate virtual experiences, fostering stronger emotional connections and enhancing the realism of social interactions. For example, in telepresence systems, high-fidelity avatars accurately capture and transmit nonverbal communication cues including facial expressions, body language, and eye contact to ensure more engaging and effective interpersonal interactions  \cite{XandraEveryBody,XandraEveryBody10,XandraEveryBody8,XandraEveryBody9}.

%Despite these advancements, current avatar creation techniques face notable challenges in generalizability, cost-effectiveness, and functionality \cite{MahrukhFromLab,highcost,huang2024wildavatarwebscaleinthewildvideo}. Many systems rely on expensive, studio-based setups and focus primarily on aesthetic personalization, such as clothing and hairstyles \cite{Xandra's82,Xandra's83}, while neglecting critical features like age-specific behaviors, diagnosis-specific physical symptoms, or nuanced emotional expressions that foster deeper connections. 

Avatar creation can be approached either manually or automatically, with each method presenting its own strengths and limitations. For manual creation, the state-of-the-art tool MetaHuman \cite{EpicGames2021, wang2023rodin} provides basic human models that require expert 3D artists to refine details such as body shapes, clothing, and facial features. Although these models are well-suited for 3D environment development, the process is both labor-intensive and time-consuming. The generated avatars are often not satisfactory to participants at first and require multiple rounds of interviews and customization before meeting their expectations \cite{MahrukhFromLab}. Furthermore, the customization options are limited, complicating the creation of highly detailed or unique avatars.

Researchers have increasingly turned to automatic avatar creation using video or image inputs to address those limitations \cite{AvatarSurvey}. NeRF-based pipelines have achieved significant progress in recent years, allowing for the automatic generation of realistic 3D human models. The introduction of 3D Gaussian Splatting (3DGS) \cite{GS} in 2023 marked a breakthrough, delivering substantial improvements in both quality and training time compared to NeRF-based methods. 

Despite these advancements, several challenges remain in adapting 3DGS-based avatars for practical VR and AR applications. Current approaches often rely heavily on public datasets, especially SMPL-X parameters, such as PeopleSnapshot \cite{peoplesnapshot} and X-Human \cite{shen2023xavatar}, which are designed to provide clean, high-quality video and images for research purposes. However, these datasets do not reflect the conditions of real-world video capture, as they rely on controlled environments with expensive equipment and labor-intensive processing that are not easily replicated in practical applications. These datasets also require extensive preprocessing to customize avatars for real-world applications, limiting their scalability and accessibility. Moreover, many systems depend on complex laboratory setups, including multiple cameras to capture 360-degree views, which increases the computational cost and demands high-performance GPUs. Additionally, while body key points are a major focus, facial expressions are often neglected, limiting the versatility of these avatars in emotionally driven applications. Furthermore, the 3DGS-based avatars are not fully integrated into real-time VR/AR platforms. They are primarily optimized for image rendering rather than interactive environments..

To address these challenges, we present an end-to-end 3DGS avatar creation pipeline designed to bridge these gaps. Our full system are publicly available at https://github.com/VU-RASL/GSAC, with corresponding Docker environment provided, and our contributions are summaried as follows:
1) We propose, to our knowledge, the first open source end-to-end pipeline for Gaussian Splatting-based avatar creation that needs only a single monocular phone-recorded video and provides a photorealistic avatar composed of a minimal amount of gaussian splats in under 40 minutes.
2) We introduce an estimation strategy for 2D missing joints using velocity over frames that will benefit optimization of  SMPL-X parameters.
3) We develop a Unity-based Gaussian Splatting Avatar Editor, allowing users to visualize and animate avatars with both Unity’s default animation system and custom joint input. We demonstrate that our pipeline produces photorealistic avatars capable of novel pose animation, offering a practical foundation for VR/AR application development.

\section{Related Work}

\subsection{From Nerf to Nerf based avatar}
Automatical avatar creation techniques can be retrived from the method of 3D recontruction. Neural Radiance Fields (NeRf) \cite{nerf}  was considered  as the state-of-the-art for 3D scene reconstruction.  It models a 3D scene as a continuous function that maps spatial coordinates and viewing directions to corresponding color and density values. By leveraging a neural network trained on input images, NeRF can generate novel views of the scene by integrating sampled rays through the 3D space. However, since NeRF represents each voxel as part of a Multi-Layer Perceptron (MLP) that requires extensive sampling and optimization, the training and rendering processes are computationally intensive and time-consuming although several researchers have  worked on reducing computation process \cite{mipnerf,mipnerf360}. 

NeRF-based photorealistic avatars,  Instant Avatar \cite{InstantAvatar}, can generate avatars from a monocular video  by leveraging human segmentation techniques like Segment Anything \cite{SAM} and body keypoint detection methods \cite{openpose,openpose1,openpose2,openpose3}.  The introduction of the Skinned Multi-Person Linear model (SMPL) \cite{SMPL} and its extended version, SMPL-X \cite{SMPLX}, enables these avatars to be trained in a canonical space and animated effectively through skeletal pose deformation. This also inspires recent methods on Gaussian Splatting Avatar creation. However, NeRF-based methods are often computationaly expensive since it needs to train a large netural works and solve rendering due to their volumetric representation, and they produce unexpected artifacts during rendering for unseen regions\cite{InstantAvatar}. Additionally, their reliance on large neural network models and the significant computational cost required for rendering each frame in real time make them impractical for deployment on VR/AR devices, where speed and scalability are critical.

\subsection{From Gaussians Splatting to Gaussian Splatting Avatar}
To address the complexity of neural network training and rendering in NeRF-based methods, Gaussian Splatting \cite{GS} was introduced. Unlike volumetric representations, it employs a point cloud-based approach, where each point in 3D space is represented as a trainable 3D Gaussian. These Gaussians efficiently capture fine details such as position, scale, rotation, and color, enabling both high-quality scene reconstruction and significantly faster training and rendering compared to NeRF-based methods. However, Gaussian Splatting originally assumes the scene is static and requires input data captured by a moving camera, which is typically preprocessed using Structure-from-Motion (SfM) methods like COLMAP \cite{sfm, sfm1}.

Yet GS-based avatar creation has faced significant challenges, particularly when using setups where the camera remains static while the human subject rotates to capture a full set of views. Achieving consistent results in such setups is difficult, as it is nearly impossible for a person to maintain a perfectly static pose. Micro-movements of both the subject and the camera introduce subtle inconsistencies, introducing blur and visual artifacts into the resulting full-view reconstructions.

Several methods have attempted to address these limitations. COLMAP-free techniques, such as the method proposed by Fu et al. \cite{Fu_2024_CVPR}, eliminate reliance on COLMAP for view alignment. Other approaches designed for dynamic scene capture, such as  \cite{TiNeuVox,4dWU,luiten2023dynamic} , aim to handle motion more effectively. However, these methods often result in significantly longer training times or require expensive laboratory camera setups. Furthermore, they are not specifically tailored for avatar reconstruction, rendering, or animation, leaving a gap in achieving practical and efficient solutions.

To overcome these challenges, state-of-the-art methods in Gaussian Splatting (GS) avatars assume dynamic input and deform the Gaussian splats for each frame using human parametric models such as SMPL \cite{SMPL} and SMPL-X \cite{SMPLX}, and Liner Skinning Blind (LBS). SMPL-X offers additional control over finger joints and facial expressions, making it particularly suitable for realistic avatar creation. Thanks to the team of SMPL-X \cite{SMPLX}, Several rigged Unity SMPL-X meshs are available to the public.  Parameters for these models can be extracted from images using methods like SMPLify-X \cite{HAHA46} or modern feed-forward approaches \cite{HAHA34,HAHA54}, enabling more accurate and adaptable avatar reconstructions.

Several researchers \cite{D3GA,jena2023splatarmorarticulatedgaussiansplatting,lei2023gartgaussianarticulatedtemplate, qian20233dgsavatar, ExAvatar, HAHA} have made significantly excellent progresses in creating photorealistic avatars using monocular video and human parametric models. Most approaches start with a canonical posed avatar, which is deformed for each frame to the target pose using SMPL or SMPL-X parameters. By incorporating Multi-Layer Perceptrons (MLPs) during training, these methods are able to capture more detailed features. For instance,  ExAvatar \cite{ExAvatar} made a significant contribution by introducing a learnable triplane architecture that enhances Gaussian features through multiple MLPs via per-vertex Gaussian asset regression. Their method achieved impressive performance, surpassing prior  approaches such as \cite{qian20233dgsavatar,hu2024gaussianavatarrealistichumanavatar}, particularly in producing photorealistic avatars with expressive facial and hand movements. They also developed a customized preprocessing pipeline and optimized SMPL-X parameters to further improve avatar quality.

Among recent advancements, HAHA \cite{HAHA} stands out as an exceptional and highly impactful work. To address the issue of excessive Gaussian counts, HAHA proposed a hydra model that elegantly combines learnable textures for SMPL-X meshes with surrounding Gaussians, resulting in visually stunning and realistic avatars. By intelligently filtering out redundant Gaussians, they successfully reduced the final count to around 10k, significantly enhancing rendering efficiency without compromising quality. Moreover, their parameterized learning framework, which binds Gaussians to the polygons of the SMPL-X mesh, provides powerful inspiration for future methods that aim to manipulate Gaussians based on mesh topology and deformation. HAHA’s approach represents a remarkable step forward in the field and has set a high standard for future research.

\begin{table}[h!]
\centering
\caption{Comparison of State-of-the-Art Full Body Avatar Methods with Our Approach.}
\label{tab:comparison_SOTA}
\resizebox{\columnwidth}{!}{%
\begin{tabular}{@{}l|p{1cm}|p{1.2cm}|p{1cm}|p{1cm}|p{1.4cm}@{}}
\toprule
\textbf{Method}         & \textbf{Custom Dataset} & \textbf{Monocular Input} & \textbf{Fast Training ($\leq$1hr)} & \textbf{Face and Hand Details} & \textbf{VR/AR Development Supports } \\ \midrule
\text{D3GA\cite{D3GA}}           &                       &                                   &                              &                                       &                                      \\
\text{3DGS Avatar\cite{qian20233dgsavatar}}    &                       & \checkmark                        &            \checkmark                   &                                       &                                      \\
\text{Gaussian Avatar\cite{hu2024gaussianavatarrealistichumanavatar}} &                     & \checkmark                        &                              &                                       &                                      \\
\text{ExAvatar\cite{ExAvatar}}       & \checkmark            &       \checkmark                              &                              &           \checkmark                              &                                      \\
\text{HAHA\cite{HAHA}}           &                       & \checkmark                        &                     \checkmark           & \checkmark                           &                                      \\
\textbf{Ours}           & \checkmark            & \checkmark                        & \checkmark                   &               \checkmark                          & \checkmark                          \\ \bottomrule
\end{tabular}
}
\end{table}

However, as shown in Table~\ref{tab:comparison_SOTA}, despite these substantial achievements, current methods still leave room for improvement when it comes to fully supporting real-time VR/AR applications. In particular, they often lack support for customizable datasets, efficient training pipelines, seamless integration with VR/AR development environments, detailed modeling of facial expressions and hand gestures, or a comprehensive combination of these capabilities.

In summary, while significant advancements have been made in creating photorealistic avatars using Gaussian splatting and human parametric models, existing methods fall short of addressing the specific needs of VR/AR applications. The limitations in dataset customization, training efficiency, VR/AR compatibility, and detailed face and hand modeling highlight the gaps in current approaches. To address these challenges, we propose a novel end-to-end Gaussian Splatting (GS) avatar creation pipeline designed to meet the demands of VR/AR development.

\section{Method}

\begin{figure*}[htbp]
\centering
\includegraphics[width=1\textwidth]{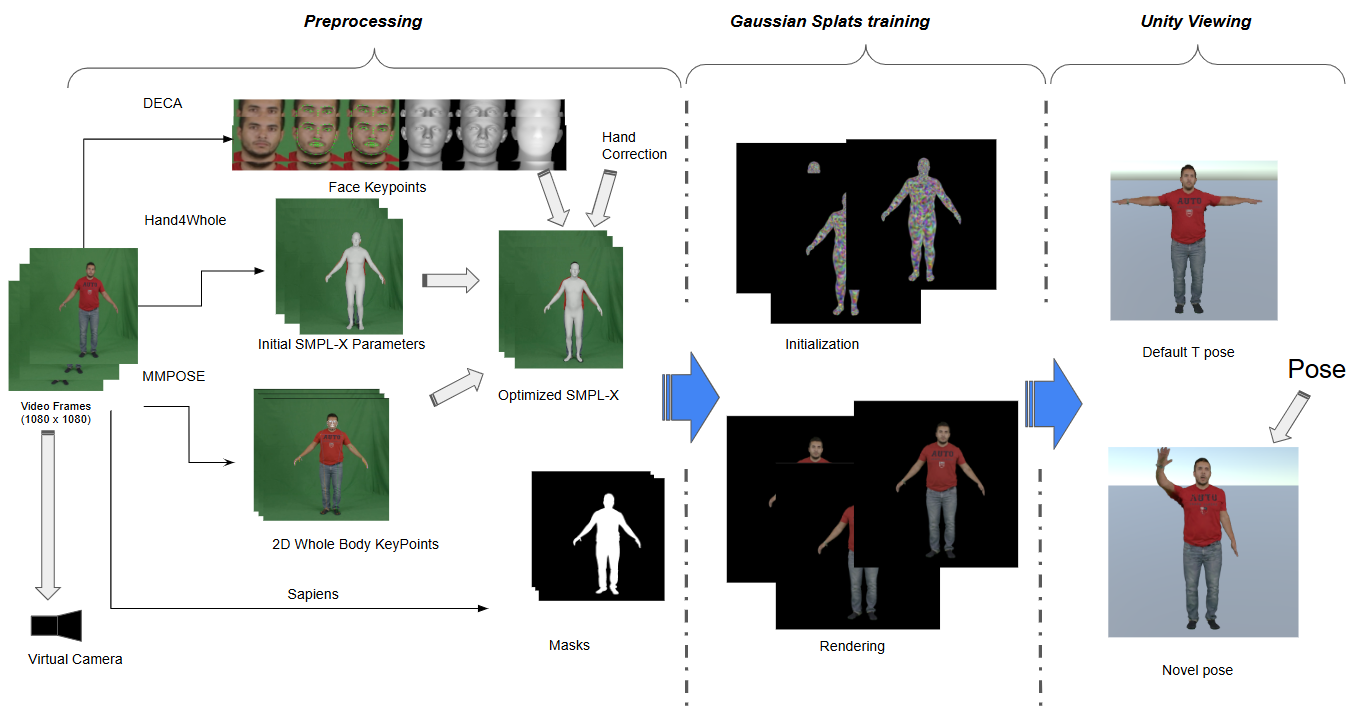} % Adjust width as needed
\caption{Overview of the proposed method, which consists of three main stages: (1) Data Preprocessing – A 1080 × 1080 frame is processed using state-of-the-art models to optimize SMPL-X parameters. (2) Gaussian Splats Training – The top row illustrates the initialization of Gaussians, while the bottom row shows the rendered images during training. For each frame, first deform the gaussian splats based on SMPL-X parameter, and then rasterized gaussians to get rendered images. (3) Unity Editor Viewing – The trained avatar is initially in a T-pose, with an a pose option provided for user-driven animation.}
\label{fig:method}
\end{figure*}

Our full pipeline (Figure~\ref{fig:method}) comprises three major components: custom data preprocessing, avatar training, and then a final Unity-based viewer. 
\subsection{Preprocessing}
To begin, we require the user to record a video where the camera remains relatively stable, and the human subject rotates within the frame. Currently, our approach supports one subject per video. The recorded video is then converted into individual frames using ffmpeg \cite{ffmpeg}. Inspired by Moon et al. \cite{ExAvatar}, we developed a data preprocessing pipeline to fit SMPL-X parameters to each input image, incorporating their optimization steps to improve parameter convergence.

First, we utilize the Detailed Expression Capture and Animation (DECA) model \cite{DECA} to extract initial FLAME (Faces Learned with an Articulated Model and Expressions) \cite{Flame} parameters, providing detailed facial geometry when faces are detected in the frames. For body modeling, we employ the Hand4Whole model \cite{Hand4Whole} to estimate SMPL-X parameters, while keypoints for the body, face, and hands are detected using the mmpose framework \cite{mmpose2020}.

We observe that when a hand is missing from the camera's view, the hand keypoints predicted by MMPose \cite{mmpose2020} are placed at the other observed hand position. This misplacement affects SMPL-X parameter optimization, as the mesh attempts to align more joint points, leading to unintended distortions. In such cases, a sudden change in body pose occurs when transitioning from a state where one hand is missing to a state where both hands are detected. We also note that MMPose performs well in predicting both shoulder positions even when one hand is absent from the camera view. While various machine learning approaches exist for pose estimation, we aim to avoid training a new network since our primary goal is to obtain a rough estimate of the missing hand position to improve the initialization of SMPL-X parameters. The SMPL-X parameters will undergo further refinement in later training stages. Therefore, we utilize a 2D kinematic chain \cite{lynch2017modern} and angular velocity to make a coarse prediction of the missing hand position.

We use the confidence scores returned by MMPose and apply a threshold to detect if a hand is missing within the frame. Specifically, if both the wrist and palm keypoint confidence scores are below the threshold, we consider the hand to be missing. We then annotate each frame in the sequence to indicate whether a hand is missing and, if so, which hand is absent.

Next, we identify the last frame before the hand disappears, denoted as \( t - n \), and the first frame where the hand reappears, denoted as \( t \). We compute the angular change between these frames for different segments of the arm: Shoulder to Elbow \( \Delta \theta_{SE}' \) , Elbow to Wrist \( \Delta \theta_{EW}' \)  and Wrist to Palm  \( \Delta \theta_{WP}' \).

Using the properties of the kinematic chain, the actual angular changes can be computed as follows:

\begin{equation}
\Delta \theta_{SE} = \Delta \theta_{SE}'
\end{equation}
\begin{equation}
\Delta \theta_{EW} = \Delta \theta_{EW}' - \Delta \theta_{SE}
\end{equation}
\begin{equation}
\Delta \theta_{WP} = \Delta \theta_{WP}' - \Delta \theta_{EW} - \Delta \theta_{SE}
\end{equation}
Finally, we compute the angular velocity for each segment over \( n \) frames:

\begin{equation}
\omega_{SE} = \frac{\Delta \theta_{SE}}{n}, \quad \omega_{EW} = \frac{\Delta \theta_{EW}}{n}, \quad \omega_{WP} = \frac{\Delta \theta_{WP}}{n}
\end{equation}
We also compute \( L_{SE}, L_{EW}, L_{WP} \) to represent the bone lengths from the \( t-n \) frame. We then update the keypoints for the missing frames sequentially (frame by frame) using the following equations:
\begin{equation}
E_{x}^{i} = S_{x}^{i} + L_{SE}^{i-1} \cdot \cos\left(\theta_{SE}^{i-1} + \Delta T \cdot \omega_{SE} \right)
\end{equation}
\begin{equation}
E_{y}^{i} = S_{y}^{i} + L_{SE}^{i-1} \cdot \sin\left(\theta_{SE}^{i-1} + \Delta T \cdot \omega_{SE} \right)
\end{equation}
\begin{equation}
W_{x}^{i} = E_{x}^{i} + L_{EW}^{i-1} \cdot \cos\left(\theta_{EW}^{i-1} + \Delta T \cdot(\omega_{SE} + \omega_{EW}) \right ) 
\end{equation}
\begin{equation}
W_{y}^{i} = E_{y}^{i} + L_{EW}^{i-1} \cdot \sin\left(\theta_{EW}^{i-1} + \Delta T \cdot(\omega_{SE}  + \omega_{EW} )\right ) 
\end{equation}
% \begin{align}
% W_{y}^{i} = E_{y}^{i} - L_{EW}^{i-1} \cdot 
% \sin\Big(& \theta_{EW}^{i-1} + V_{SE} \cdot \Delta T \notag \\
%         &+ V_{EW} \cdot \Delta T \Big)
% \end{align}
\begin{equation}
P_{x}^{i} = W_{x}^{i} + L_{WP}^{i-1} \cdot \cos\left(\theta_{WP}^{i-1} + \Delta T \cdot (\omega_{SE}  + \omega_{EW}  + \omega_{WP}) \right  )
\end{equation}
\begin{equation}
P_{y}^{i} = W_{y}^{i} + L_{WP}^{i-1} \cdot \sin\left(\theta_{WP}^{i-1} + \Delta T \cdot(\omega_{SE}  + \omega_{EW} + \omega_{WP} ) \right  )
\end{equation}
where \( i \) is the frame number ranging from \( t-n+1 \) to \( t-1 \). $\Delta T $ equals to $i - t + n$. Here, \( L_{SE}^{i-1} \) represents the bone length from  shoulder to  elbow in the previous frame; \( L_{EW}^{i-1} \) represents the bone length from  elbow to  wrist in the previous frame; \( L_{WP}^{i-1} \) represents the bone length from wrist to  palm in the previous frame, and \( \theta_{SE}^{i-1} \) is the shoulder-to-elbow angle from the previous frame, \(\theta_{EW}^{i-1} \) is the elbow to wrist angle from the previous frame,  \(\theta_{WP}^{i-1} \) is the wrist to palm angle from the previous frame. With this analysis we can estimate the new missing joint position.

Once the keypoints are identified, we segment the foreground and background using Sapiens: Foundation for Human Vision Models \cite{sapiens} depth model.

To achieve accurate alignment, we utilize a co-registration approach for SMPL-X optimization. This process refines the SMPL-X parameters by incorporating both 2D keypoints and 3D mesh constraints. Specifically, we minimize the total loss $L_{\text{SMPLX}}$ formulated as:
\begin{equation}
    L_{\text{SMPLX}} = L_{\text{kpt}} + 0.1 L_{\text{init}} + L_{\text{face}} + L_{\text{reg}},
\end{equation}
where $L_{\text{kpt}}$  represents the L1 distance of 2D keypoints detected by MMPOSE \cite{mmpose2020}and projected keypoints from the initialization of regressed parameters from \cite{Hand4Whole}.  $L_{\text{init}}$ is the L1 distance between the SMPL-X parameters being optimized and
the initial SMPL-X. 

The face loss, $L_{\text{face}}$, is defined as:
\begin{equation}
    L_{\text{face}} = 10L_{\text{vertex}} + 10000L_{\text{lap}} + L_{\text{edge}},
\end{equation}
where $L_{\text{vertex}}$ is the L1 distance between the face vertices and FLAME  vertices from DECA\cite{DECA}, $L_{\text{lap}}$ is the L2 distance to minimize Laplacian differences between SMPL-X face mesh and FLAME mesh, and $L_{\text{edge}}$ is the L1 distance that  enforces edge length regularization between SMPL-X face mesh and FLAME mesh. It is computed only when the face is visible, determined by the angle of the face relative to the camera. The visibility is assessed by calculating the dot product of two vectors: one from the face center to the midpoint of the eyes and the other from the camera's position to the face center in the $xz$-plane of the camera-centered coordinate system. The face is considered visible if the angle between these vectors is greater than $135^\circ$ (i.e., the dot product is less than $\cos(135^\circ)$).  $L_{\text{face}}$ is only computed when face is visable in the frame.

The regularization term, $L_{\text{reg}}$, is defined as:
\begin{equation}
    L_{\text{reg}} = 0.01L_{\text{shape}} + 100L_{jo} + L_{\text{sym}},
\end{equation}
where $L_{\text{shape}}$ is a squared L2 norm body shapes parameters in SMPL-X, $L_{jo}$ is a squared L2 norm of the joint offset being added to T-Pose template mesh to deform the mesh to prevent an extreme joint
offset suddenly and $L_{\text{sym}}$ is to regularize non-visible face parts, proposed by \cite{symloss}.

$L_{jo}$ is a squared L2 norm of the joint offset, which prevents an extreme joint
offset, and Lsym is for symmetricity of the joint offset and face offset, similar to
Feng et al. [12].

Camera parameters for each frame are initialized with dummy values: no rotation, no translation, and a centered camera with approximated focal lengths. These initializations serve as a baseline for further processing.

\subsection{Avatar Training}
To train the Gaussian splatting avatar, we adopted  the foundational first stage training framework introduced by HAHA \cite{HAHA} and applied it to our preprocessed dataset. This framework was chosen for its ability to optimize SMPL-X parameters while training and using Gaussian splatting for photorealistic rendering, as well as its fast training time. Moreover, their method achieves state-of-the-art quality while significantly reducing the number of Gaussians compared to other approaches. HAHA's Gaussian splatting deformation method improves efficiency and accuracy by associating each Gaussian with an SMPL-X \cite{SMPLX} mesh polygon and deforming it based on mesh movement, outperforming traditional Linear Blending Skinning (LBS) for Gaussian deformation. This approach which allows animation using reduced number of splats and achieves high quality results,  provides an opportunity to animate the Gaussians in VR/AR environments due to the limitation of computational power of VR/AR hardware. 

While training, the objective is on optimizing local Gaussian transformations: position (\( \mu\)), rotation (\( r\)), (scale \(s\)) and colors (\( c \)).  The gaussians are initalizalized at the center of each polygon in SMPL-X mesh. For a given frame with SMPL-X parameters, the final transformations are defined relative to the SMPL-X mesh polygons:
\begin{equation}
    \mu' = kR\mu + T
    \label{eq:mu_update}
\end{equation}
\begin{equation}
    r' = Rr
    \label{eq:r_update}
\end{equation}
\begin{equation}
s' = ks
\label{eq:s_update}
\end{equation}

Here, \( \mu' \) represents the Gaussian's translation, calculated as an offset to the polygon center, where \( k \) is the scale of polygon, \( R \) is the quaternion
rotation of the polygon, and \( T \) is the translation of the polygon center. \( r' \) represents the Gaussian's rotation derived directly from the polygon's rotation, and \( s' \) represents the Gaussian's scale, computed relative to the polygon's scale. Colors (\( c_i^r \)) is RGB colors. The body pose and body shape parameters are treated as the learnable parameters as well.
The total loss for training process is formulated as:

\begin{equation}
L_{\text{Gaussian}} = L_2 + 0.01 L_{\text{LPIPS}} + 0.1 L_{\text{SSIM}} + L_{\text{Sobel}} + 0.01 L_{\text{KNN}}
\end{equation}

where \( L_2 \) loss measures pixel-wise differences, \( L_{\text{LPIPS}} \)\cite{lpips} is a perceptual loss that captures high-level visual similarities, \( L_{\text{SSIM}} \) is a structural similarity loss for preserving image structure, \( L_{\text{Sobel}} \) computes the sharpness difference using the \( L_2 \) distance between Sobel-operator \cite{soberOperator} outputs on rendered and ground-truth images, and \( L_{\text{KNN}} \) is a regularization term that ensures smoothness by minimizing the standard deviation of Gaussian properties among neighboring Gaussians, which is introduced by Lei et al. 
 \cite{lei2023gartgaussianarticulatedtemplate}.

\subsection{Unity Editor }
Building on the foundation established by Pranckevičius\cite{UnityGaussianSplattingToy}, we developed an enhanced editor for Gaussian Splatting in Unity. Gaussian splats are rendered as 2D quads on the GPU, linearly blended each frame's traditionally rasterized background based on alpha value. In regions where the destination alpha is high (more opaque), the contribution of new splats is diminished, whereas in more transparent areas, they appear more prominently. The destination color (previously rendered content) remains fully intact, facilitating smooth composition. Additionally, a fragment shader applies an exponential falloff based on the Gaussian’s position, meaning the further away a part is, the more transparent it becomes, further refining transparency to create a natural blending effect.

 In addition, we integrated the publicly available rigged SMPL-X mesh \cite{SMPLX} to animate the trained Gaussian splats using user-specified pose parameters. To facilitate animation, we created a custom compute shader updates Gaussian splats properties based on changes in the SMPL-X mesh. When a new pose is provided, the mesh vertices are updated, and the corresponding Gaussian transformations, as described in Equations~\eqref{eq:mu_update} \eqref{eq:r_update} \eqref{eq:s_update}, are computed within the compute shader. Since the SMPL-X mesh supports Unity default animation, our Gaussian Avatar supports it as well. These updated properties are then stored in the GPU buffer, ensuring efficient real-time rendering of the animated Gaussian splats.   To the best of our knowledge, this is the first Unity program that supports animatable Gaussian Avatar with Unity default animation.

\subsection{Experimental Design}

Since the preprocessing pipeline mainly focus on estimating and optimizing SMPL-X parameters captured from raw video, we only evaluate the generated SMPL-X parameters.  We assume there is only one person in the video, and we assume the gender is known for that person (male or female).

\textbf{Dataset} PeopleSnapShot :  This dataset, published by \cite{peoplesnapshot}, meets our requirement since we assume the desired input involves a human rotating in place with a static camera. Additionally, their unmasked raw inputs are publicly available. We selected four subsets from the dataset: female-3-casual, female-4-casual, male-3-casual, and male-4-casual.

HAHA utilized the improved SMPL parameters estimated by AnimNerf \cite{animnerf} from the original PeopleSnapshot frames and further refined hand and face expressions using SMPLify-X \cite{smplifyx} to create the preprocessed dataset, as AnimNerf \cite{animnerf} only provides SMPL \cite{SMPL} parameters. 

For a fair comparison, we directly adopt HAHA's \cite{HAHA} original experimental settings. We maintain the random color background during training to prevent Gaussians from learning the background. The same training and testing frames are processed to estimate SMPL-X parameters. To ensure consistent ground truth for comparison, we use the same masks provided in HAHA's preprocessed dataset. The ground truth images are background removed images by the masks. We train and test Gaussian avatars three times: one using our preprocessed SMPL-X estimation without hand correction, one with hand correction and one using HAHA's preprocessed SMPL-X estimation. We then compute PSNR, SSIM, and LPIPS as evaluation metrics by running each dataset five times and averaging the results.  Note that AnimNeRF optimizes the SMPL mesh parameters during training, with the entire training process taking approximately 13 hours on two RTX 3090 GPUs.\textit{ In contrast, our custom preprocessing for SMPL-X estimation takes roughly 20–35 minutes using a single RTX 4090 GPU for approximately 80–120 frames.}

\section{Results}

Our contribution is in two parts: 1. the preprocessing pipeline for customizing dataset, and 2) Unity editor for viewing and animating Gaussian splattings. We will evaluate the result in two sections.
\subsection{Preprocessing Pipeline}

Table ~\ref{tab:comparison_male} and Table ~\ref{tab:comparison_female} show that our method significantly reduces training time while maintaining comparable or better performance in several aspects compared to HAHA. Across all subjects, our full pipeline consistently requires less training time, with an average reduction of 230.1 seconds $\pm $ 3.41 seconds.

The total number of Gaussians generated by our method is comparable to that of HAHA, often resulting in slightly fewer Gaussians.

In terms of quality metrics, our method achieves competitive results: Without hand correction, the PSNR and SSIM scores are generally slightly lower than those of HAHA. However, after applying our hand correction strategy, both PSNR and SSIM improve significantly, demonstrating the effectiveness of this correction. For \textit{Male4} and \textit{Female3}, our full pipeline often outperforms the ground truth. In the case of \textit{Female4}, the PSNR is approximately 0.13 lower than the ground truth, and the LPIPS is higher by 0.002, though the SSIM is slightly improved.

\begin{table*}[h!]
\centering
\caption{Comparison of training time, number of Gaussian splats (GS), PSNR, SSIM, and LPIPS for male subjects (casual) using HAHA's preprocess and our method.}
\label{tab:comparison_male}
\begin{adjustbox}{max width=\textwidth}
\begin{tabular}{@{}lcccccccccc@{}}
\toprule
  & \multicolumn{3}{c}{Male3} & \multicolumn{3}{c}{Male4} \\ 
\cmidrule(lr){2-4} \cmidrule(lr){5-7}
 & HAHA preprocess  & GSAC (ours) preprocess & GSAC (ours w. hand correction) preprocess & HAHA preprocess & GSAC (ours) preprocess  & GSAC (ours w. hand correction) preprocess\\ \midrule

Training time (secs) 
& 947.24 $\pm$ 4.83 
& 806.22 $\pm$ 6.44 
& \cellcolor{yellow} 716.41$\pm$ 9.10 
& 931.38 $\pm$ 5.75 
& 791.06 $\pm$ 5.60 
&  \cellcolor{yellow} 696.78 $\pm$ 4.62\\

Number of GS 
& 25996  $\pm$ 65 
& 25416  $\pm$ 153 
& \cellcolor{yellow} 25347.2  $\pm$  134
& 28372  $\pm$ 51 
& 28015  $\pm$ 59 
& \cellcolor{yellow} 27986.8  $\pm$  95\\

PSNR 
& \cellcolor{yellow}30.6862  $\pm$ 0.0120 
& 28.6914  $\pm$ 0.0079 
& 29.7016 $\pm$ 0.0344
& 26.5799  $\pm$ 0.0364 
& 26.3338  $\pm$ 0.0344 
&  \cellcolor{yellow} 26.8484 $\pm$ 0.0197 

\\

SSIM 
& \cellcolor{yellow}0.9612  $\pm$ 0.0002 
& 0.9493  $\pm$ 0.0001 
& 0.9543  $\pm$ 0.0001 
& 0.9600  $\pm$ 0.0001 
& 0.9579   $\pm$ 0.0001 
& \cellcolor{yellow}0.9606  $\pm$ 0.0001 
\\

LPIPS 
&\cellcolor{yellow} 0.0433  $\pm$ 0.0004 
& 0.0568  $\pm$ 0.0003 
& 0.0552  $\pm$ 0.0001
& \cellcolor{yellow}0.0488  $\pm$ 0.0002 
& 0.0552  $\pm$ 0.0001 
& 0.0544  $\pm$ 0.0003 
\\ 
\bottomrule
\end{tabular}
\end{adjustbox}
\end{table*}

\begin{table*}[h!]
\centering
\caption{Comparison of training time, number of Gaussian splats (GS), PSNR, SSIM, and LPIPS for female subjects (casual) using HAHA's preprocess and our method.}
\label{tab:comparison_female}
\begin{adjustbox}{max width=\textwidth}
\begin{tabular}{@{}lcccccccccc@{}}
\toprule
  & \multicolumn{3}{c}{Female3} & \multicolumn{3}{c}{Female4} \\ 
\cmidrule(lr){2-4} \cmidrule(lr){5-7}
 & HAHA preprocess  & GSAC (ours) preprocess & GSAC (ours w. hand correction) preprocess & HAHA preprocess & GSAC (ours) preprocess  & GSAC (ours w. hand correction) preprocess\\ \midrule
Training time (secs) &   941.15  $\pm$ 4.23 & 799.57  $\pm$ 4.37 & \cellcolor{yellow} 709.20 $\pm$ 7.52 & 939.79  $\pm$ 2.35 & 773.488  $\pm$ 5.04 &
\cellcolor{yellow} 713.68 $\pm$ 2.13\\

Number of GS 
&   \cellcolor{yellow}26950 $\pm$ 51 
& 27396 $\pm$ 85 
& 27218 $\pm$ 87
& 25150 $\pm$ 59 
& 25114 $\pm$ 118 
& \cellcolor{yellow}24958 $\pm$ 84

\\

PSNR 
&   28.3468 $\pm$ 0.0266 
&  29.4070 $\pm$ 0.0341  
& \cellcolor{yellow}29.9854 $\pm$ 0.0215 
& \cellcolor{yellow}29.0738 $\pm$ 0.0337 
& 28.6041 $\pm$ 0.0402 
& 28.9497 $\pm$ 0.0341  \\

SSIM 
&   0.9558 $\pm$ 0.0001 
& 0.9565 $\pm$ 0.0001 
& \cellcolor{yellow}0.9580 $\pm$ 0.0001 
& 0.9529 $\pm$ 0.0002 
& 0.9521 $\pm$ 0.0002
& \cellcolor{yellow}0.9533 $\pm$ 0.0001 

\\

LPIPS 
&   0.05928 $\pm$ 0.0003
& 0.0566 $\pm$ 0.0003
&   \cellcolor{yellow}0.0560 $\pm$ 0.0002
& \cellcolor{yellow}0.0440 $\pm$ 0.0004
& 0.0469 $\pm$ 0.0004 
& 0.0465 $\pm$ 0.0002 

\\ \bottomrule
\end{tabular}
\end{adjustbox}
\end{table*}

\begin{figure*}[!htbp]
    \centering
    \includegraphics[width=0.95\textwidth]{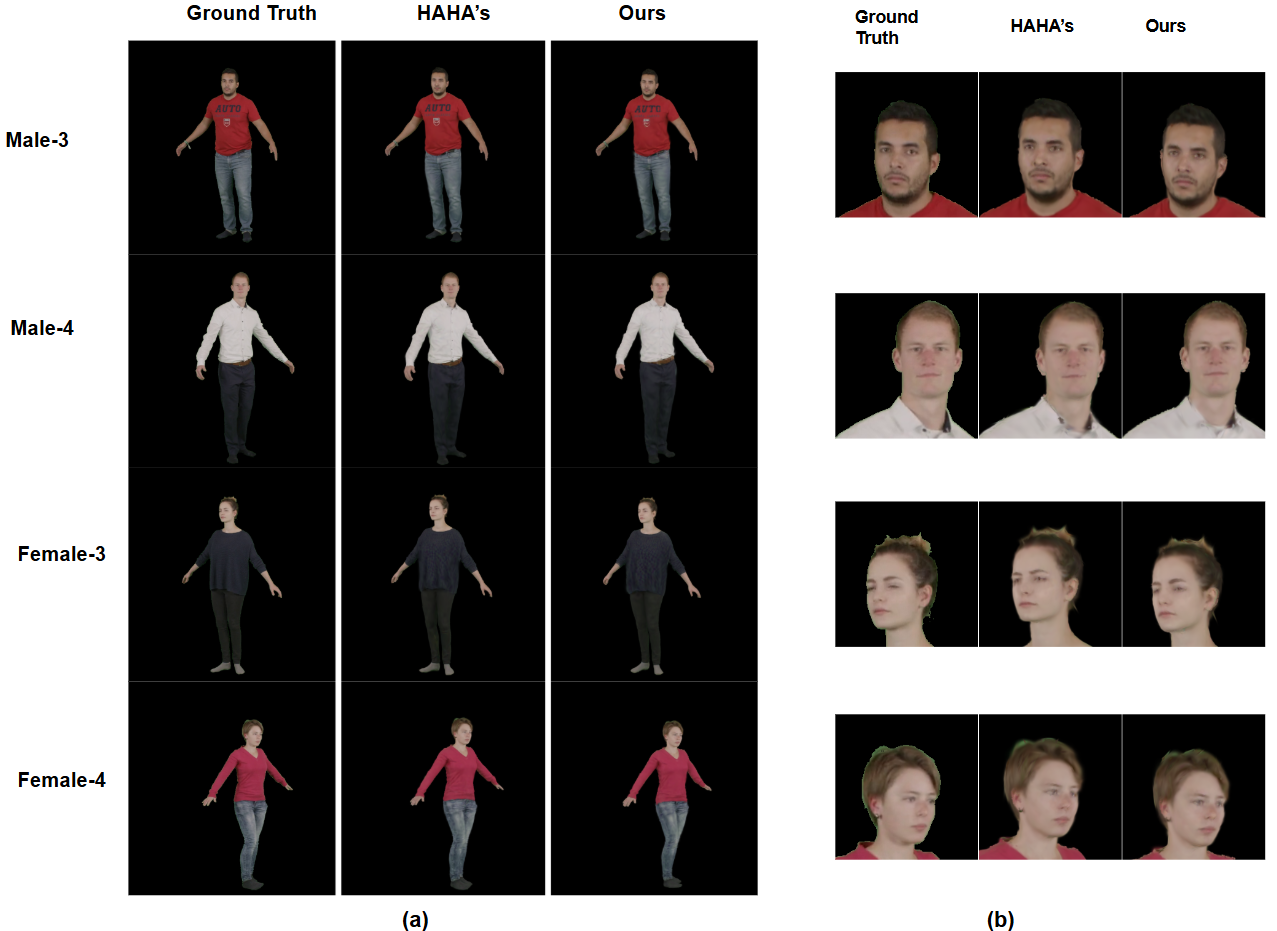} 
    \caption{Comparison of rendered results for each subject. (a) Full-body visualizations show that both our method and HAHA produce shapes closely matching the ground truth. (b) Cropped views of facial expressions demonstrate that our method achieves higher fidelity in capturing facial details compared to HAHA.}
    \label{fig:qualresult_peoplesnapshot}
\end{figure*}

In Figure~\ref{fig:qualresult_peoplesnapshot}, we present the visualized results for each subject, by rendering its test dataset. The full-body visualizations demonstrate that both HAHA's and our method produce body shapes that closely resemble the ground truth images. In Figure~\ref{fig:qualresult_peoplesnapshot}(b), we highlight facial expressions by cropping the same area in each image. As shown in the results, our method captures facial expressions with greater accuracy compared to HAHA's.

\begin{figure*}[htbp]
    \centering
    \includegraphics[width=0.95\textwidth]{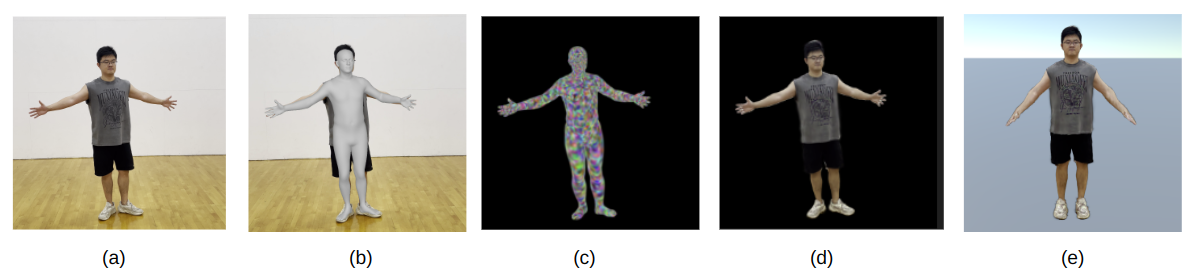}
    \caption{ Qualitative results on a volunteer's video input, illustrating key steps of our pipeline. (a) Resized image frame(1080x1080) from input frame captured using an iPhone 12 Pro (1440 × 1440 resolution). (b) Visualization of estimated SMPL-X parameters. (c) Initialized Gaussians (d) Gaussian rendering while training (e) Final trained Gaussian splats avatar in Unity, presented in an A-pose.}
    \label{fig:qualresult_customdata}
\end{figure*}

We also present qualitative results on a custom dataset captured by a volunteer recorded 20 seconds of video for each key step, as shown in Figure ~\ref{fig:qualresult_customdata}. In (a), we capture a frame using an iPhone 12 Pro at a resolution of 1440x1440 and resized to 1080x1080 serving as the input of our pipeline. In (b), we fitted SMPL-X parameters of that frame and we visualize the estimated SMPL-X parameters. In (c), we show the initialization of gaussians on the center of each polygons of the mesh. In (d), we illustrate the Gaussian training process and showcase its rendered image. Finally, in (e), we present the trained Gaussian splats avatar in Unity, posed in an A-pose.

\subsection{Unity Editor}

\begin{figure*}[!htbp]
    \centering
    \includegraphics[width=0.9\textwidth]{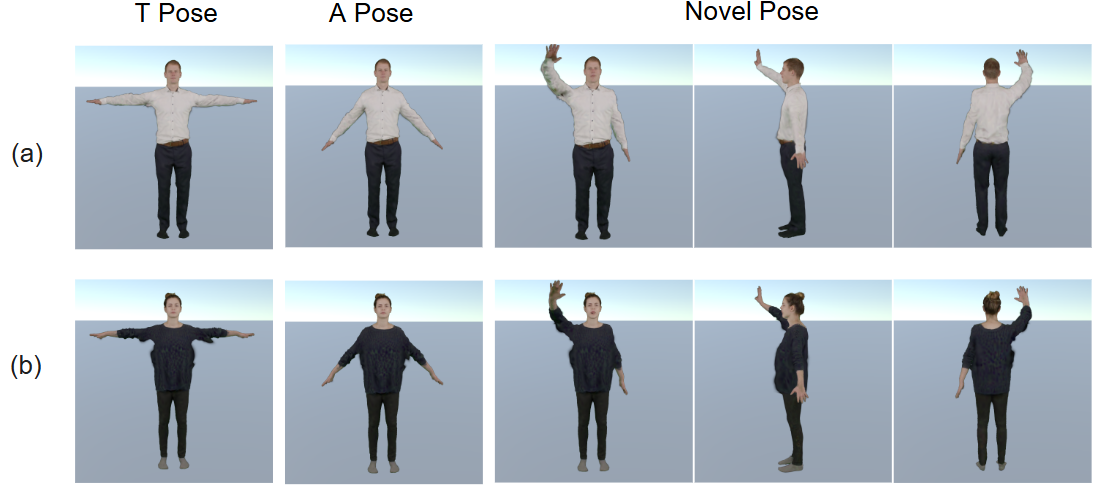} 
    \caption{Qualitative results of our Unity viewer, demonstrating three different poses—T-pose, A-pose, and a novel pose with a simple facial expression—for (a) male4 and (b) female3 from the PeopleSnapshot dataset.}
    \label{fig:unityview}
\end{figure*}

Secondly, we qualitatively demonstrate our Unity viewer. We first show that our Gaussian avatar supports customized joint rotations. Figure~\ref{fig:unityview} showcases three different poses— a T-pose, an A-pose, and a novel pose with a simple facial expressionfor \textit{male4} and \textit{female3} from the PeopleSnapshot dataset. The novel pose is defined using our pose controller, which takes the position of each joint as input. Each avatar is fitted with SMPL-X, trained using our pipeline, and subsequently imported into Unity.

\begin{figure*}[!htbp]
    \centering
    \includegraphics[width=0.9\textwidth]{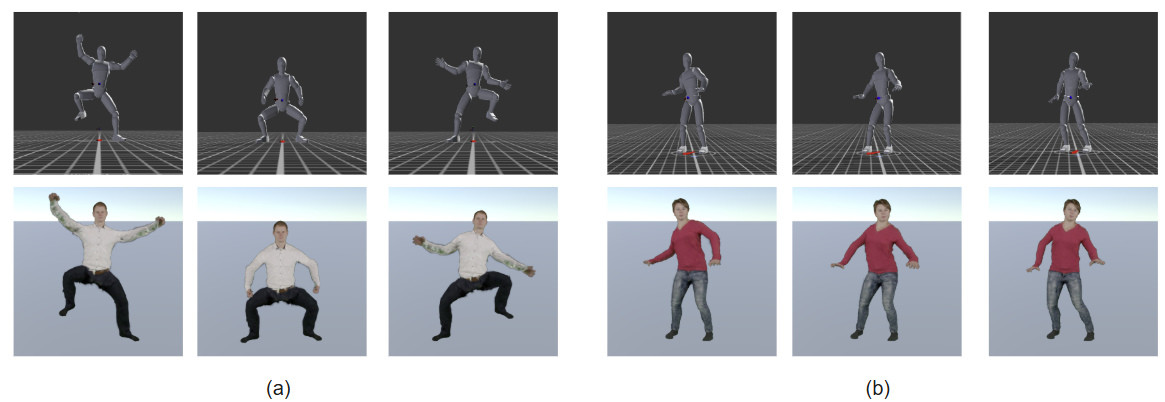} 
\caption{Demonstration of Gaussian avatars animated with Unity’s default animation system. (a) Avatar of Male4 animated with selected frames of dance animation, obtained from the Unity Asset Store. (b) Avatar of Female4 animated with selected frames of another dance animation}

    \label{fig:unityview1}
\end{figure*}

Figure~\ref{fig:unityview1} further demonstrates that our Gaussian avatars are compatible with Unity's default animation system. In subfigure (a), we apply a dance animation, obtained from the Unity Asset Store \cite{unity_dance_animations}, and capture 3 keyframes. Subfigures (b) is animated female4 with another dance animation from the same package

% \begin{table}[htbp]
% \centering
% \caption{Average FPS for each Avatar.}
% \begin{tabular}{@{}llllll@{}}
% \toprule
%         &  Male3 &  Male4 &  Female3 &  Female4 &  Custom \\ \midrule
% Avg FPS &    $74.04 \pm $8.79   &   $69.18 \pm $7.94    &    $70.43 \pm $8.86     &    $76.71 \pm $10.41     &    $65.35 \pm $8.48    \\ \bottomrule
% \label{tab:fps}
% \end{tabular}
% \end{table}

For the novel pose, we also provide side and back views to further illustrate the model's fidelity. As shown in the figure, the trained avatars exhibit high-quality rendering and can be animated accurately based on the given pose. However, some artifacts remain. Certain regions, such as the area beneath the arms, remain ambiguous or under-reconstructed due to their absence in the input frames. Additionally, Gaussian points that were not deformed or observed during training tend to exhibit inaccurate placements or rotations when subjected to novel poses. This limitation arises from the nature of the image-based supervision used in Gaussian Splatting: the model can only learn from what is visible in the training video frames. As a result, unseen regions receive little to no supervision, leading to potential artifacts. 

We update the avatar's pose every 0.5 seconds, cycling between the three poses shown in Figure~\ref{fig:unityview}, and measure the frames per second (FPS) every second. Each avatar runs for 10 minutes, and we compute the final average FPS. Male3, Male4, Female3, and Female4 have FPS of  $74.04 \pm 8.79 $, $69.18 \pm 7.94$, $70.43 \pm 8.86$ , $76.71 \pm 10.41$,  respectively. All avatars achieve an FPS above 60, ensuring smooth and fast rendering.

\section{Discussions}

In this work, we introduced a novel pipeline for generating photorealistic avatars using Gaussian splatting, seamlessly integrated into Unity for potential VR/AR applications. Our results demonstrate that the proposed pipeline achieves competitive avatar quality, faster training times, and fewer Gaussian splats while maintaining a quality comparable to HAHA. 

\subsection{Limitations}
Due to the nature of Gaussian splatting, which relies on image loss during training, reconstructing unseen parts of the human body remains challenging. For instance, underarm details are difficult to capture when the arms are not fully visible as the person rotates. 

Our current Gaussian splats model is trained for 3,000 iterations, producing fair-quality results. However, further exploration is needed to determine the optimal number of splats required to balance quality and efficiency. As training iterations increase, more Gaussians are added to capture finer details, but it is essential to investigate the trade-off between rendering quality and computational cost.

Since this work does not focus on optimizing mask generation, some artifacts may appear in the personalized avatars, particularly when using the custom dataset.
These artifacts are most noticeable in complex regions such as hair, where segmentation inaccuracies can affect the final visual quality.

Currently, clothing and garments are trained as an integral part of the body surface, meaning any outfit change requires recording a new video and retraining the avatar. While this approach produces visually consistent results, it limits user flexibility. A more convenient solution would involve enabling direct customization or modular swapping of clothing on the avatar. Additionally, the current pipeline lacks support for realistic cloth dynamics, as garments do not exhibit independent physical movement or deformation in response to body motion.

\subsection{Future Works}

To address the challenge of reconstructing unseen regions of the human body, future work could explore the use of generative models to synthesize novel views based on the observed parts. By leveraging learned priors about human shape, pose, and appearance, these models could hallucinate plausible geometry and texture for occluded or unobserved areas. This would enable the use of video fragmented or limited views of a human to generate a dataset of input frames with a sufficient amount of views for high-quality avatar creation using GSAC.

Since Sapiens \cite{sapiens} provides both full-body masks and accurate body-part segmentation, we can refine training to exclude specific clothing regions—for example, training an avatar without the upper-body clothing area. This would enable a separate clothing model to be trained and bound to the corresponding mesh faces, allowing for interchangeable outfits without requiring additional video recordings. As Gaussians are bound to the underlying mesh, the system does not capture physical dynamics such as cloth motion or deformation, as garments move rigidly with the body mesh.

Future research could focus on extending the pipeline to support real-time animation, dynamic user interaction, and broader integration within VR/AR applications. In particular, by leveraging our custom pose controller, it would be possible to connect real-time motion capture (MoCap) streams to the Gaussian-based avatar. This would enable responsive, real-time interaction within immersive environments, further bridging the gap between photorealistic Gaussian avatars and fully interactive virtual experiences.

With continued advancements, more avatars can be created efficiently and made interactive in real time, benefiting a wide range of application areas. One compelling example is simulation-based training in healthcare, where lifelike avatars help overcome challenges in creating realistic and diverse training scenarios. Traditional methods often rely on physical manikins or standardized patients (SPs) \cite{Xandra5,Xandra6}, trained individuals who simulate real-life clinical conditions across different demographics. However, SPs come with high operational costs, ethical concerns, and logistical limitations—especially in sensitive areas like pediatric care \cite{Xandra12,Xandra16}. Photorealistic, dynamic avatars offer a scalable alternative, enabling learners to practice technical procedures, clinical decision-making, and inter-professional collaboration in psychologically safe and repeatable virtual environments.

\section{Conclusion}
In this work, we present the first end-to-end pipeline for creating a Gaussian Splatting-based avatar using only a phone-recorded video in approximately 40 minutes. Our approach leverages several cutting-edge machine learning models and introduces novel techniques for optimizing SMPL-X parameters. The resulting parameters yield visually compelling and often superior results compared to existing methods. The system not only supports Unity’s default animation system but also allows users to input custom SMPL-X parameters, enabling personalized or motion-captured animations based on real human motion. This makes the avatar readily usable in VR/AR applications with minimal additional setup.

While there remains room for improvement, this work lays a strong foundation for the next generation of immersive VR/AR experiences—demonstrating that realistic, animatable avatars can be generated quickly and effectively with accessible tools.

\section{Acknowledgement}

The human images used in this paper are sourced from the publicly available PeopleSnapshot~\cite{peoplesnapshot} dataset as well as from our custom dataset, which was collected with the assistance of volunteers. We gratefully acknowledge the authors  for making their dataset publicly available. We also extend our sincere appreciation to all the volunteers who participated in our data collection process. Written consent was obtained from each volunteer for the use of their images in this work.

We sincerely appreciate the support provided for this research by the National Institutes of Health (NIH) under award R21AG078480 and the National Science Foundation (NSF) under award 2225890. Their funding has been instrumental in advancing our work.

\bibliography{references}
\bibliographystyle{IEEEtran}

\end{document}